\documentclass[12pt]{iopart}
\usepackage[dvips]{graphics}
\usepackage{graphicx}
\usepackage{iopams}
\begin{document}
\title[Phase ordering and universality for continuous symmetry models on graphs]{Phase ordering and universality for continuous symmetry models on graphs}
\author{R Burioni$^1$, F Corberi$^2$, A Vezzani$^3$}
\address{$^1$ Dipartimento di Fisica and INFN, Universit\`a di Parma, 
Parco Area delle Scienze 7/A, 42100 Parma, Italy}
\address{$^2$ Dipartimento di Matematica ed Informatica
via Ponte don Melillo, Universit\`a di Salerno, 84084 Fisciano (SA), Italy}
\address{$^3$ CNR-INFM, S3, National Research Center, via Campi 213/a, 41100 Modena, Italy and  Dipartimento di Fisica, Universit\`a di Parma, 
Parco Area delle Scienze 7/A, 42100 Parma, Italy}

\begin{abstract}

We study the phase-ordering kinetics following a temperature quench
of $O(\cal N)$ continuous symmetry models with ${\cal N}=3$ and ${\cal N}=4$
on graphs.
By means of extensive simulations, we show that the global pattern 
of scaling behaviors is analogous to the
one found on usual lattices.
The exponent $a_\chi$ for the integrated response function
and the exponent $z$, describing the growing length, are related to the
large scale topology of the networks through the spectral dimension and the fractal
dimension alone, by means of the same expressions 
provided by
the analytic solution of the $\cal N\to \infty $ limit.
This suggests that
the large-$N$ value of these exponents 
could be exact for every ${\cal N}\geq 2$.
\end{abstract}


\maketitle

\def\be{\begin{equation}}
\def\ee{\end{equation}}
\def\bfi{\begin{figure}}
\def\efi{\end{figure}}
\def\bea{\begin{eqnarray}}
\def\eea{\end{eqnarray}}

\section{Introduction}

A ferromagnetic system quenched from a high temperature disordered state to
an ordered phase with broken ergodicity evolves via phase-ordering
dynamics.
In the  late stage of growth,  the correlation of the order parameter
between sites $i,j$ at times $s,t$ can be expressed as the
sum of two terms \cite{Furukawa89}
\be
{\cal C}_{ij}(t,s)={\cal C}^{st}_{ij}(t-s)+{\cal C}^{ag}_{ij}(t,s).
\label{1}
\ee
The first term describes the
contribution provided by degrees of freedom which
in the interval $[s,t]$ are not interested by non-equilibrium
effects, such as the passage of topological defects, 
while the second contains the non-equilibrium
information.
Analogously, also the integrated response function, or zero field
cooled magnetization,  measured on site $i$
at time $t$
after a perturbation has been switched on in $j$ from time $s$ onwards,
takes an analogous addictive form~\cite{bouchaud}
\be
\chi_{ij}(t,s)=\chi^{st}_{ij}(t-s)+\chi^{ag}_{ij}(t,s).
\label{1a}
\ee
On regular lattices, due to space homogeneity and isotropy,
correlation and response function depend only
on the distance $r$ between $i$ and $j$. One has, therefore,
${\cal C}_{ij}(t,s)={\cal C}(r,t,s)$, and similarly for $\chi_{ij}(t,s)$.

The non-equilibrium behaviour is characterized by the
dynamical scaling symmetry, a self-similarity where
time acts as a length rescaling. When scaling
holds, the states sequentially probed by the system are statistically
equivalent provided lengths are measured in units of
the characteristic length scale $L(t)$, which increases in time.
All the time dependence enters through $L(t)$,
and the aging parts in Eqs.~(\ref{1},\ref{1a}) take a scaling form
in terms of rescaled variables~\cite{Furukawa89}
$x=r/L(s)$ and $y=L(t)/L(s)$
\be
{\cal C}^{ag}(r,t,s)=\tilde {\cal C}(x,y),
\label{2}
\ee
\be
\chi ^{ag}(r,t,s)=s^{-a_\chi}\tilde \chi (x,y).
\label{2a}
\ee
The characteristic length usually grows according to a power law 
\be
L(t)\sim t^{1/z}.
\label{scallen}
\ee
Interestingly, non-equilibrium exponents are expected
to be universal, namely to depend only on a restricted set of parameters.
On regular
lattices, where a substantial understanding of the dynamics has been
achieved by means of exact solutions, approximate theories
and numerical simulations~\cite{Bray94}, these 
exponents depend only on
the space dimensionality, the number of components ${\cal N}$ of the order
parameter and the conservation
laws of the dynamics.
In this paper we will consider primarily $z$ and $a_\chi $.
It is known~\cite{Bray94}, that $z$ depends only on the conservation
laws, being $z=2$ in the case of a non conserved order parameter
considered in this paper. 
Regarding $a_\chi$, for continuous symmetry models (${\cal N}>1$) with 
non-conserved order parameter, it was conjectured in ~\cite{noi}
to obey
\be
    a_{\chi} = \left \{ \begin{array}{ll}
        \frac {d-2}{2}  \qquad $for$ \qquad d < 4  \\
        1  \qquad $with log corrections for$ \qquad d=4 \\
	1   \qquad $for$ \qquad d > 4. 
        \end{array}
        \right .
        \label{expa}
\ee
Let us  notice that these exponents share the property of being 
${\cal N}$-independent.
Their value therefore can be computed in the soluble large-${\cal N}$ model~\cite{noininf}. 
The reference framework of a soluble theory is a great advantage in the study of 
phase-ordering on networks or inhomogeneous
graphs considered in this paper. In fact our understanding of these systems 
is largely incomplete, although examples can
be found in
disordered materials, percolation clusters, glasses, polymers,
and bio-molecules, and are also present in interdisciplinary studies,
ranging from economics to chemistry and social sciences~\cite{interd}.
The aim of this paper is to investigate if also on networks 
the non-equilibrium dynamics of statistical models
takes a scaling structure, and its  universality features. 
Moreover, it is interesting
to understand which
topological indices of the graph play the role of the euclidean dimension $d$
on regular lattices in determining universal quantities.
We will restricted our attention to {\it physical graphs} \cite{rassegna}: 
These are networks with the appropriate topological features to
represent real physical structures, namely they  are embeddable in a finite
dimensional space and have bounded degree. 

Equilibrium properties of models defined on  physical
graphs, and in particular the relevance of their 
topology, are quite well understood.
As far as systems with continuous symmetry are concerned, such as 
$O({\cal N})$ models (${\cal N}\geq 2$), a 
unique parameter,
the "spectral dimension" $d_s$, encodes the relevant large scale
topological features of the network and regulates the
critical properties. $d_s$ is related to the low
eigenvalues behaviour of the density of state of the
Laplacian operator \cite{spectral}, and can be considered
for these models as the topological indicator replacing the
Euclidean dimension $d$ on graphs:
The spectral dimension univocally determines the existence of phase
transitions
\cite{fss} and controls critical behaviours \cite{sferico}, much in the
same way as the Euclidean dimension $d$ does on usual translation
invariant lattices.

Regarding non-equilibrium, in the large-${\cal N}$ model it was 
shown~\cite{Bettolo97,prlnostro} that 
the general framework of
scaling behaviour discussed above is maintained on generic graphs.
In particular the topology of the network enters the exponent $z$
only through the fractal dimension $d_f$ and
the spectral dimension $d_s$, 
\be
z=2d_f/d_s,
\label{zinfty}
\ee
while $a_\chi$ depends only on the spectral dimension obeying~(\ref{expa}) 
with $d_s$ occurring in place of $d$.

In this paper we want to complement the large-${\cal N}$ analysis by studying
the phase-ordering kinetics
of some $O({\cal N})$ vector models with finite ${\cal N }$ 
on a class of  physical graphs.
We will consider geometrical fractals without phase transition
at finite temperature, such as the Sierpinski
gasket or the T-fractal, and others
structures, obtained by direct products among graphs \cite{rassegna},
which on the contrary feature a phase
transitions at a finite temperature $T_c$.
Our simulations of these systems evolving with relaxational dynamics 
(non-conserved order parameter)
show that, for ${\cal N}=3$ and  ${\cal N}=4$, 
the dynamical exponents $z$ and $a_\chi $ 
are correctly predicted by the large-${\cal N}$ model  
and then depend only on the fractal 
and spectral dimensions  $d_f, d_s$.
This suggests that
in presence of scaling these exponents take the same value 
for any ${\cal N}\geq 2$, much in the same way as it
happens on euclidean lattices.

This paper is organized as follows: In Sec. \ref{Physical Graphs} we 
introduce the physical graphs giving a definition of fractal and spectral 
dimension.
In Sec.~\ref{model} we define the
$O({\cal N})$ models that will be considered in the simulations. We also introduce
the basic observables, and discuss the numerical techniques.
In Sec. \ref{Results} we present our results for
different structures.
Sec.~\ref{concl} contains a final discussion and the conclusions.

\section{Physical Graphs} \label{Physical Graphs}

A graph (network) $\cal G$ is a discrete structure defined by a set of sites $i$ connected pairwise by unoriented links (edges) $\{i,j\}$. The chemical distance $r_{i,j}$, i.e. the number of links in the shortest path connecting  sites $i$ and $j$, naturally defines on $\cal G$ a metric. Van Hove spheres , allowing to explore large  scales of the graph, can be constructed using this metric. The van Hove sphere ${\cal S}_{o,r}$ of radius $r$ and center $o$ is the sub-graph of $\cal G$ composed by the sites whose distance  from $o$ is smaller than $r+1$. Calling $N_{o,r}$ the number of sites in ${\cal S}_{o,r}$, the fractal dimension of the graph is defined by its asymptotic behaviour at large scales
\be
N_{o,r}\sim r^{d_f}
\label{fractal_dim}
\ee
where  $\sim$ denotes the behaviour for large $r$. In the following we will consider only physical graphs embeddable in a finite dimensional space ($d_f$ is well defined and finite). Moreover we require that the degree $z_i$ (number of neighbours of the site $i$) is bounded.

Algebraic graph theory provides powerful tools for the description of the topology of 
generic networks, by means of characteristic matrices. 
The adjacency matrix $A_{i,j}$ of a graph has entries equal to $1$ if $i$ 
and $j$ are neighbouring sites ($\{i,j\}$  is a link) and $A_{i,j}=0$ otherwise. 
The Laplacian matrix $\Delta_{i,j}$ is defined as
\be
\Delta_{i,j}=\delta_{i,j} z_{i}-A_{i,j}
\label{laplacian}
\ee    
where $z_i=\sum_j A_{i,j} $ is the degree of $i$. Interestingly, $\Delta_{i,j}$ is the generalization to graphs of the usual Laplacian operator of Euclidean structures \cite{rassegna}. In particular its spectrum is positive and the constant vector is the only eigenvector of eigenvalue zero. Moreover, on physical graphs the spectral density  $\rho(l)$ of $\Delta_{i,j}$, is expected to behave as \cite{spectral}
\be
\rho(l)\sim l^{d_s/2-1}
\label{spectral_dim}
\ee 
where $\sim$ denotes the behaviour for small $l$'s. Eq. (\ref{spectral_dim}) defines the spectral dimension $d_s$ of the physical graph.

In order to build graphs of larger dimensions we introduce the direct product between graphs. 
Given $\cal G$ and $\cal H$ the direct product ${\cal G}\times {\cal H}$ is a graph whose 
sites are labelled by a pair $(i,j)$ with $i$ and $j$ belonging to $\cal G$ and $\cal H$ 
respectively. $(i,j)$ and $(i',j')$ are neighbour sites in ${\cal G}\times {\cal H}$ 
if $i=i'$ and $\{j,j'\}$ is a link of $\cal H$, or if $j=j'$ and $\{i,i'\}$ is a link of 
$\cal G$. A basic properties of ${\cal G}\times {\cal H}$ is that, calling  $d_s^{\cal G}$ 
and $d_f^{\cal G}$ the dimensions of the graph $\cal {G}$, one has \cite{rassegna}
\be
d_s^{{\cal G} \times {\cal H}} = d_s^{{\cal G}}+d_s^{{\cal H}}~~~~~~~
d_f^{{\cal G} \times {\cal H}} = d_f^{{\cal G}}+d_f^{{\cal H}}~~~~~~~
\label{sum_spectral_dim}
\ee 
i.e. the spectral and fractal dimensions of the product graph are the sum of the dimensions of the original graphs.

In the simulations,  we will consider models defined on graphs with known $d_s$ and $d_f$ 
in order to verify the relevance of these dimensions, describing the large scale 
topology, on phase ordering. 
In particular, we focus on two finitely ramified fractals \cite{aharony} with $d_s<2$, the T-fractal and the Sierpinski gasket, whose fractal and spectral dimensions can be analytically evaluated by means of exact renormalizations \cite{rammal}, yielding $d_f=\log(3)/\log(2)$, $d_s=\log(9)/\log(6)$ and $d_f=\log(3)/\log(2)$, $d_s=\log(9)/\log(5)$ respectively. Moreover, in order to explore the region between $d=2$ and $d=3$,  we also consider the graphs obtained from the product of two gaskets and two T-fractals, which both have spectral dimensions $2< d_s< 3$ .

\begin{figure}
    \centering
\rotatebox{0}{\resizebox{.5\textwidth}{!}{\includegraphics{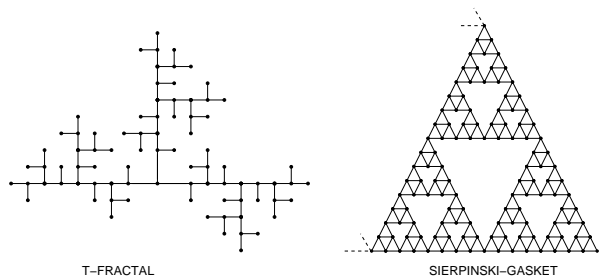}}}
      \caption{The finitely ramified fractals considered in the paper.}
\label{strutture}
\end{figure}

\section{The Model} \label{model}

The Hamiltonian of  $O({\cal N})$ spin systems on a graph is given by:
\be
H_0[\vec \sigma ]=-J\sum _{i,j }^NA_{i,j} \vec \sigma _i \cdot 
\vec \sigma _j
\label{hamiltonian}
\ee                  
where $|\vec \sigma _i| = 1$ are unitary $\cal N$-component vectors,  
and $N$ is the number of sites in the network. In the following we will
set $J=1$. 
Phase ordering is obtained by evolving an initially fully disordered 
configuration with a dynamics at a temperature $T$ where the system presents 
an ordered phase at equilibrium. In the following we will consider 
quenches to $T=0$. This guarantees that the dynamics is of the
phase-ordering type even for those systems with $d_s\leq 2$,
namely without phase transition at finite temperature.

Phase ordering can be characterized by the
correlation function, defined as 
\be
{\cal C}_{ij}(t,s)= \langle \vec \sigma _i(t) \cdot \vec \sigma _j(s)\rangle -
\langle \vec \sigma _i(t)\rangle \cdot \langle \vec \sigma _j(s)\rangle
\label{correl}
\ee
Recalling Eqs. (\ref{1},\ref{2}), the non-equilibrium scaling properties are
encoded in the aging part ${\cal C}_{ij}^{ag}(t,s)$ which,
in turn, can be obtained by subtracting the equilibrium contribution
$C_{ij}^{st}(t,s)$ from the whole correlation ${\cal C}_{ij}(t,s)$.
Notice however that at $T=0$, ${\cal C}_{ij}^{st}(t,s)$ vanishes and  
hence ${\cal C}_{ij}^{ag}(t,s)={\cal C}_{ij}(t,s)$. We will drop therefore the superscript
$ag$ from the correlation function in the following.
In order to simplify the analysis one usually restricts the attention
to ${\cal C}_{ij}(t,t)$, namely the equal time correlation function,
or to the autocorrelation function ${\cal C}_{ii}(t,s)$.
On lattices ${\cal C}_{ij}(t,t)=G(r,t)$ is a function of the euclidean distance $r$ 
between $i$ and $j$ which, according to Eq. (\ref{2}) scales as 
\be
G(r,t)=g(r/L(t))
\label{corrlen}
\ee
This allows one to extract $L(t)$ as, for instance, the half-height
width of $G(r,t)$. 
Concerning the autocorrelation function, on lattices
it does not depend on $i$, due to translational invariance,
so that ${\cal C}_{ii}(t,s)=C(t,s)$ with, following Eq. (\ref{2}), the scaling form 
\be
C(t,s)=f(y)\sim y^{-\lambda},
\label{scalauto}
\ee
where $\lambda$ depends both on $d$ and on $\cal N$ \cite{Bray94}. 

When phase-ordering occurs on graphs one has the additional
feature of the dependence of ${\cal C}_{ij}(t,t)$ on
$i$ and $j$ (and ${\cal C}_{ii}(t,s)$ on $i$), complicating the analysis and hindering
the scaling properties. Therefore, 
in order to simplify the study, we will resort in the following to 
particular correlations, with a transparent physical meaning, which where
shown in~\cite{prlnostro,lungonostro}
to be useful in detecting scaling properties.
More precisely, for the equal time correlation function,
on the Sierpinski gasket and on the T-fractal
we will compute it restricting $i$ and $j$ on the
baseline of the structure (see Fig. \ref{strutture}), as this procedure allows to
soften the log-periodic oscillations characteristic of deterministic fractals \cite{oscillations} and gives the best
results for the scaling plots, as will be shown in Sec. \ref{Results}. 
Regarding the autocorrelation
function we will consider its spatial average 
${1\over N} \sum_{i=1}^N {\cal C}_{ii}(t,s)$, in order to obtain the best statistic. For simplicity
these quantities will be denoted as $G(r,t)$ and $C(t,s)$ in analogy
to their counterparts on lattices.

The response of the system to an external field can be studied 
introducing the susceptibility
\be
\chi_{ij}(t,s)=\lim_{h\to 0}{1\over h} \langle  \vec \sigma_i (t) 
\rangle_{h_j(s)} \cdot \vec n_j 
\label{susc}
\ee
where $\langle \vec \sigma_i (t) \rangle_{h_j(s)}$ denotes the expectation value 
at time $t$ of the spin $\vec \sigma_i$ when a field $\vec h_j=h \vec n_j$
($\vec n_j$ being a unitary vector),
changing the Hamiltonian to 
$H[\vec \sigma ]=H_0[\vec \sigma ]- \vec \sigma _j \cdot\vec h_j$ is switched 
on from time $s<t$ onwards on site $j$. 
Differently from what happens for ${\cal C}_{ij}(t,s)$, the equilibrium contribution 
to $\chi_{ij}(t,s)$ does not vanish. 
Then, in order to isolate
the scaling part $\chi_{ij}^{ag}(t,s)$, according to Eq. (\ref{1a}) one has 
to subtract $\chi_{ij}^{st}(t-s)$
from the whole response measured during the quench.  
$\chi_{ij}^{st}(t-s)$ can be numerically evaluated as the response
of a system prepared initially in the equilibrium state at $T=0$,
namely with all the spins aligned. Actually, if one does this
the system starts to evolve 
as soon as a small field $h>0$ is switched on even if $T=0$, 
originating a non-vanishing response. 
Considering, for simplicity, the autoresponse $\chi(t,s)=\chi_{ii}(t,s)$,
on lattices, from Eq. (\ref{2a}) one finds 
\be
\chi^{ag}(t,s)=s^{-a_\chi}h(y)
\label{scalrespag}
\ee
with $a_\chi $ depending only on $d$ according to Eq. (\ref{expa}).
On networks the response function is site dependent, so,
similarly to what done for the autocorrelation function, we will consider
the spatially averaged quantity
${1\over N} \sum_{i=1}^N \chi_{ii}(t,s)$ which will be denoted
as  $\chi (t,s)$. 

Zero temperature dynamics can be implemented in different ways. 
For example one can set $T= 0$ in a Metropolis updating rule. 
A convenient choice adopted in this paper consists in aligning a 
randomly chosen spin $\vec\sigma_i$ with the local field, 
i.e. the spin $\vec \sigma_i $ is turned into 
\be
\vec \sigma_i \to \vec \sigma_i'=|h| \vec n_i+ \sum_{j}A_{i,j} \vec\sigma_j .
\label{move}
\ee
With this choice, at  each move the local energy is minimized. 
Such a simple updating rule proves to be very efficient, on networks
and lattices as well. For instance 
it allows to 
reproduce easily the analytically known behaviour of the $O(2)$ model in
one-dimension \cite{rutenberg}, whereas the Metropolis rule fails due
to extremely long transients (see Appendix for details).  Notice that in 
one-dimensional $O(2)$ model scaling is violated, and our 
hypothesis on dynamical exponents does not hold.

The scope of this study being the analysis of the behaviour of
finite ${\cal N}$ models, the less
numerically demanding case to start with would naturally be the $O(2)$ model. 
However,  as discussed in the Appendix, the dynamics of 
this model is pinned on self-similar graphs such as the Sierpinski gasket
or the T-fractal considered here. 
Then, in order to study the phase
ordering kinetics in this case, one should allow the system to depin
by quenching to a small but finite temperature. Since the above mentioned
structures, however, are disordered at any finite temperature, in order to 
recover true phase-ordering one should then let $T\to 0$, 
in the same way as for phase-ordering in the one-dimensional
Ising model with Kawasaki dynamics \cite{cornell}. 
Since this procedure is very numerically demanding,   
hereafter we will focus on the next simplest cases, namely with ${\cal N}=3$
and ${\cal N}=4$ (the latter restricted to the Sierpinski gasket). 
Performing extensive 
numerical simulation of these models on different self-similar graphs, we show that  
phase ordering exhibits dynamical scaling, similarly to
what is known on lattices, in any case. In particular,  
the correlation length grows according to Eq. (\ref{scallen})
and the two time correlation and response functions scale as (\ref{scalauto})
and (\ref{scalrespag}).
We also find that the exponents $z$ and $a_\chi $ 
are well consistent with their large-${\cal N}$ value.
This suggests that also on graphs, as on lattices, these exponent 
do not depend on ${\cal N}$, and that their value can be
predicted by the solution of the large-${\cal N}$ model for
every scaling system with ${\cal N}\ge 2$.

\section{Results} \label{Results}

In the numerical simulations we considered  the Sierpinski gasket, the T-fractal, 
and the graphs obtained by a direct product of two gaskets and of two T-fractals. The number of sites $N$ of the structures is 2391486, 1549324,10771524 and 4787344 respectively.
A time-step is made of $N$ elementary Montecarlo moves (\ref{move}).
The data are obtained by averaging over at least 500 dynamical realizations, each
with a different random initial condition. For the calculus of $\chi(s,t)$ a 
different random field $\vec n_i$ is applied in each run.  
To evaluate  the limit $h\to 0$ in Eq. (\ref{susc}), we have simulated systems with 
small external fields $h$ and then we have verified that $\chi(t,s)$ is independent of 
$h$ (e.g. comparing systems with field $h$ and $h/2$). 
In our simulations $h$ varies between $10^{-1}$ and $10^{-3}$. 

\begin{figure}
    \centering
   \rotatebox{0}{\resizebox{.5\textwidth}{!}{\includegraphics{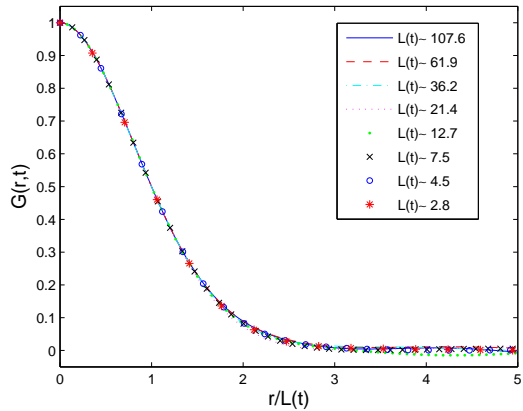}}}
    \caption{Scaling of $G(r,t)$ evaluated on the baseline of a Sierpinski gasket.}
\label{scal_corr_gas}
\end{figure}

\begin{figure}
    \centering
   \rotatebox{0}{\resizebox{.5\textwidth}{!}{\includegraphics{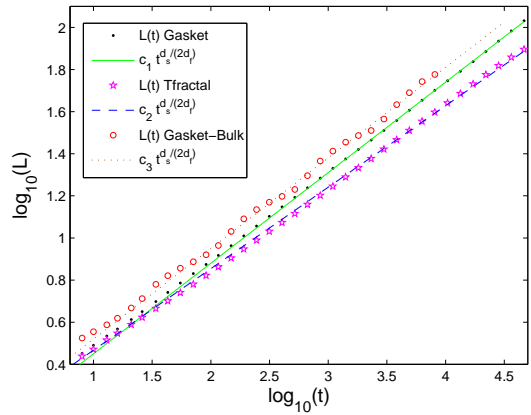}}} 
    \caption{Growth of $L(t)$ as a function of time, 
anomalous diffusive behaviour $L(t)\propto t^{1/z}$ 
is well verified. For distances evaluated along the baseline, best fits yield $1/z=0.42\pm 0.01$ and $1/z=0.39 \pm 0.01$ for the Sierpinski
gasket and the T-fractal, respectively, very well consistent with
the large-${\cal N}$ behaviour $1/z={d_s/(2d_f)}$ in both cases. 
For the Sierpinski Gasket we also plot $L(t)$, evaluated averaging over sites
belonging to the bulk of the structure. The value of the exponent 
does not change, but the log-periodic oscillations are much larger.}
\label{Lcorr_GAS_TF_BULK}
\end{figure}

In Figure \ref{scal_corr_gas} we show the behaviour of the equal time correlation
function $G(r,t)$ defined in Sec. \ref{model} for the Sierpinski gasket. 
The Figure evidences an excellent scaling, in the form expected from Eq. (\ref{corrlen}). 
Quite surprisingly, the scaling is very good even for very short times 
($t$ ranges from 8 to 46600) evidencing the efficiency of the dynamics considered. 
Analogous results were obtained for the T-fractal. 
In Figure \ref{Lcorr_GAS_TF_BULK} we plot the growth of $L(t)$ as a function of 
time, finding a very good agreement with the asymptotic behaviour $L(t)\sim t^{d_s/(2 d_f)}$
(namely $t^{0.431}$ and $t^{0.387}$ for the Sierpinski gasket and the T-fractal
respectively).
We note that superimposed to the expected behaviour there are small 
log-periodic oscillations, typical of fractal structures \cite{oscillations}.  If the average of the correlation function 
is not restricted on the baseline of the structure, but is taken
on the whole fractal, the log-periodic oscillations are more evident, as shown in Figure \ref{Lcorr_GAS_TF_BULK} for the Sierpinski gasket, while keeping the same slope for the $L(t)$.
If this average procedure is used, it is more difficult to extract the scaling behaviour of the correlation function.

\begin{figure}
    \centering
   \rotatebox{0}{\resizebox{.5\textwidth}{!}{\includegraphics{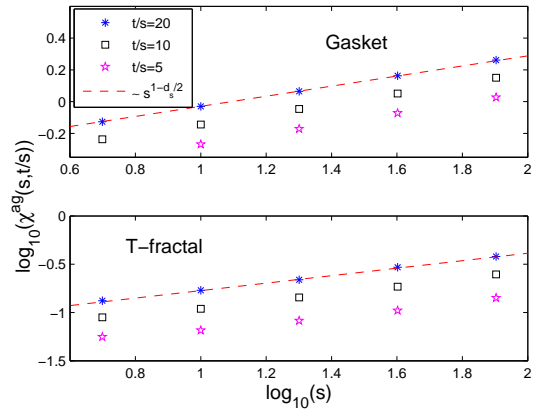}}} 
    \caption{Estimate of the exponent $a_\chi$ for the Sierpinski gasket and the T-fractal
(see text). 
Best fits yield $a_\chi=-0.32\pm 0.01$ for the Sierpinski gasket and $a_\chi=-0.38\pm 0.01$ 
for the T-fractal. 
In both cases these values are consistent with the hypothesis $a_\chi=d_s/2-1$.}
\label{scaling_chi_gas_tf}
\end{figure}

\begin{figure}
    \centering
   \rotatebox{0}{\resizebox{.5\textwidth}{!}{\includegraphics{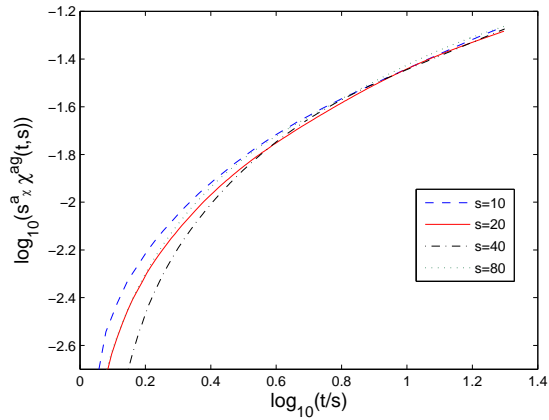}}} 
    \caption{Scaling of the function $\chi^{ag}(t,s)$ on the Sierpinski gasket. 
At large time ratio, a good agreement with the scaling hypothesis (\ref{scalrespag}) is verified for $s\geq10$.}
\label{scal_chi}
\end{figure}

\begin{figure}
    \centering
   \rotatebox{0}{\resizebox{.5\textwidth}{!}{\includegraphics{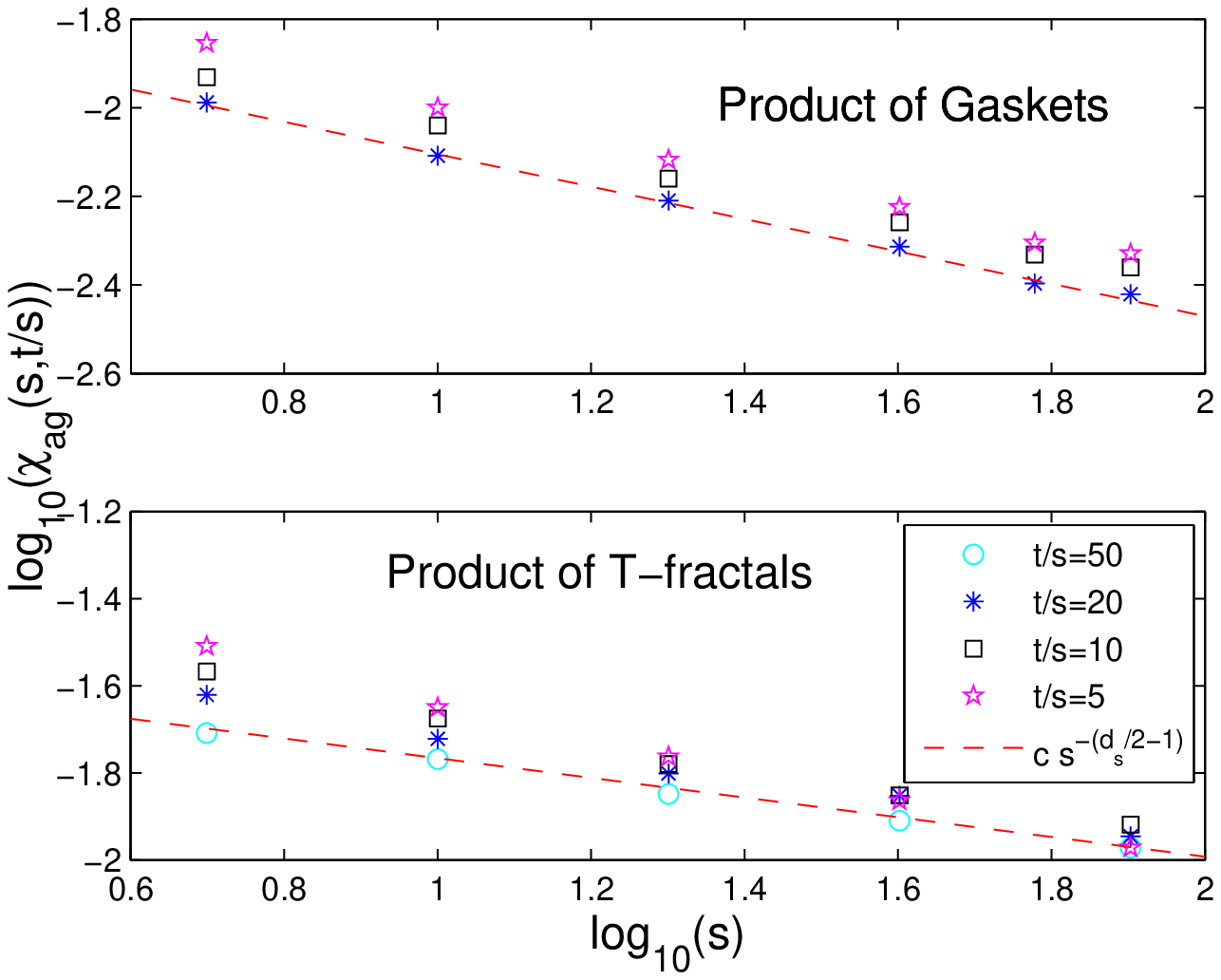}}} 
    \caption{Estimate of the exponent $a_\chi$ for the product of Sierpinski gaskets and 
the product of T-fractals. 
Best fits yield $a_\chi=0.36\pm 0.01$ for the product of Sierpinski gaskets and 
$a_\chi=0.22\pm 0.01$ for the product of T-fractals. 
In both cases a good agreement with the large-${\cal N}$ value is 
evidenced. Notice that in the first case the asymptotic regime is reached 
much  earlier 
than in the second one.}
\label{scaling_chi_gas2_tf2}
\end{figure}

Let us now turn to the response function.
It has been analyzed as follows: After computing $\chi^{ag}(t,s)$ by subtracting
$\chi^{st}(t,s)$ to the full non-equilibrium response, as explained in
Sec. \ref{model},
we fix the ratio $y=t/s$ and plot $\chi^{ag}(t,s)$ versus $s$
(Fig. \ref{scaling_chi_gas_tf}). In doing so we obtain, 
from (\ref{scalrespag}), an estimate for $a_{\chi}$ from the slope of the plot. 
As illustrated in 
Fig. \ref{scaling_chi_gas_tf}, both for the Sierpinski gasket and for 
the T-fractal the value of $a_\chi$  is consistent with $d_s/2-1$ 
(namely $a_\chi =-0.317$ for the Sierpinski gasket and $a_\chi = -0.387$
for the T-fractal)
i.e. the analytic expression obtained in the ${\cal N}\to \infty$ case
(except for the smaller values of $t/s$, as we will discuss below). 
Finally, we verified the scaling relation (\ref{scalrespag}) by plotting 
$s^{a_\chi}\chi^{ag}(t,s)$ versus $t/s$ for different values of $s$ and
checking for data collapse, as shown for the Sierpinski gasket in Fig. \ref{scal_chi}.
The collapse is good for $t/s\gtrsim 5$, whereas it is poor for smaller values
of $t/s$. This can be interpreted as due to finite-$s$ preasymptotic corrections,
which are more effective at small $t/s$, similarly to what observed on 
lattices \cite{noi,noi2}. 
Analogous result can be obtained for the T-fractal. 
Let us stress the particular feature of the cases with $d_s<2$ of 
$\chi^{st}(t-s)$ (and hence $\chi(t,s)$) scaling itself as $s^{-a_\chi}\tilde h(t/s)$,
with the same exponent $a_\chi $ of $\chi ^{ag}(t,s)$ and with a scaling function
$\tilde h(y)$ that, although different from $h(y)$, has the same
asymptotic behaviour $\tilde h(y)\simeq y^{-a_\chi}$ for $y\gg 1$.
In this case, therefore, one gets the same information on the exponents
from  $\chi (t,s)$, $\chi ^{ag}(t,s)$ and $\chi ^{st}(t-s)$. 
The situation is different for $d_s>2$. Here  the stationary term
converges to the finite equilibrium susceptibility, 
$\lim _{t\to \infty} \chi^{st}(t-s)=\chi_{eq}$, whereas $\chi ^{ag}(t,s)$ scales
with a positive $a_\chi$ (see below). 
In this case, therefore,  $\chi (t,s)$ is dominated by $\chi ^{st}(t/s)$
and the subtraction of $\chi ^{st}(t/s)$ is necessary in order to make the
scaling of the aging part manifest.
In Fig. \ref{scaling_chi_gas2_tf2} we evaluate the exponent $a_\chi$,
following the procedure discussed above, both for the product of gaskets 
and for the product of T-fractals. Also in this case we get a very good agreement 
with the expected behaviour $a_\chi=d_s/2-1$ (namely $a_\chi =0.365$ and $a_\chi=0.226$
for the product of gaskets and the product of T-fractals respectively).


\begin{figure}
    \centering
   \rotatebox{0}{\resizebox{.5\textwidth}{!}{\includegraphics{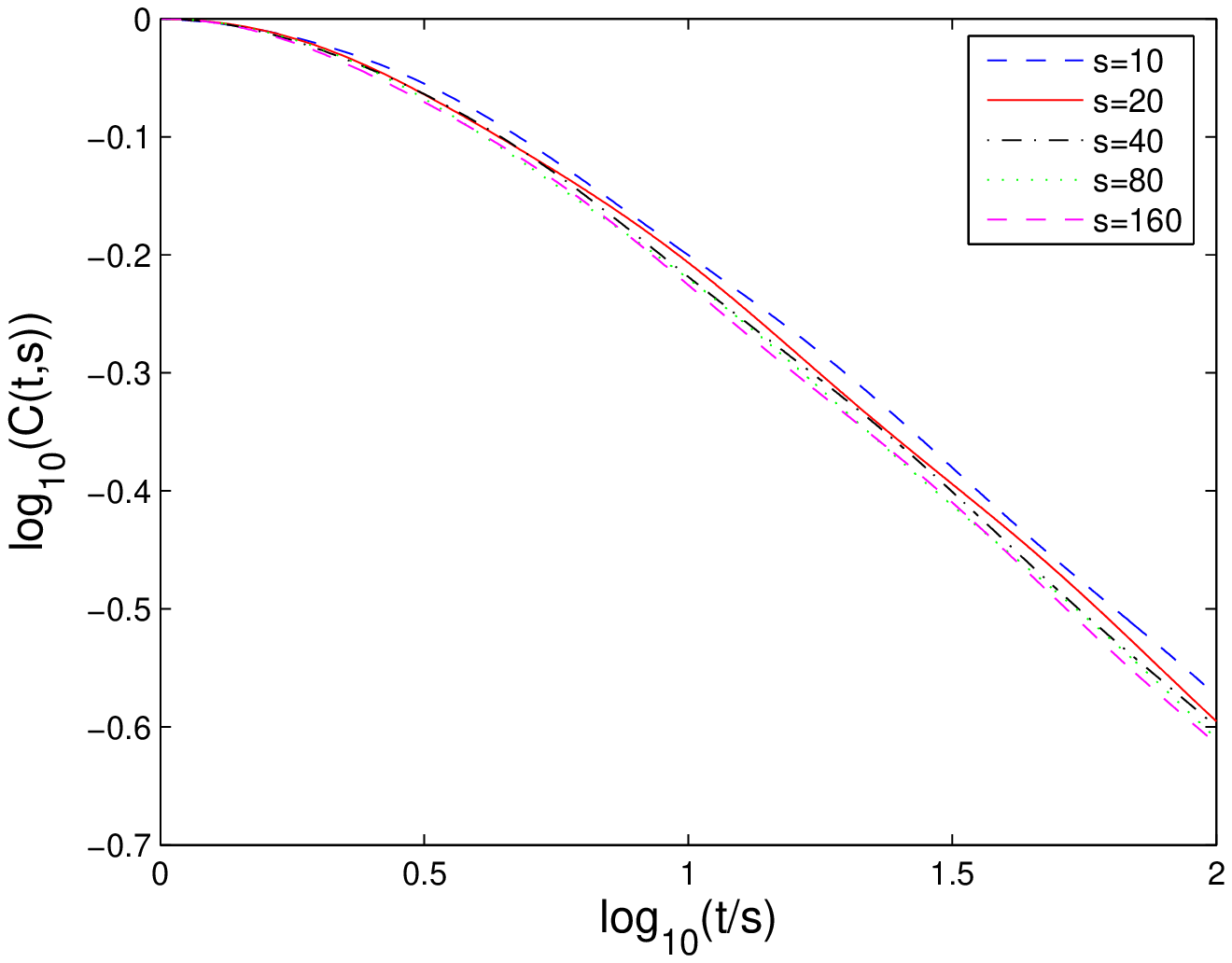}}} 
    \caption{Scaling of the two time correlation function $C(s,t)$.
A good agreement with the hypothesis $C(s,t)=h(s/t)$ is found.}
\label{scal_c_gas}
\end{figure}

In order to provide a further check on dynamical scaling,
we have considered also the two-time correlation function $C(t,s)$
introduced in Sec. \ref{model}. 
The scaling relation (\ref{scalauto}) is well verified,
as shown  in Fig. \ref{scal_c_gas} for the Sierpinski gasket.
A residual correction,
due to the finite values of $t_w$ can be observed for large $t/t_w$,
similarly to what is known in lattices \cite{Brown}.
The quality of the scaling improves by increasing $t_w$, with the curves
for $t_w=40,80,160$ almost collapsing, as expected. 
Similar results were found for the T-fractal.

\begin{figure}
    \centering
   \rotatebox{0}{\resizebox{.5\textwidth}{!}{\includegraphics{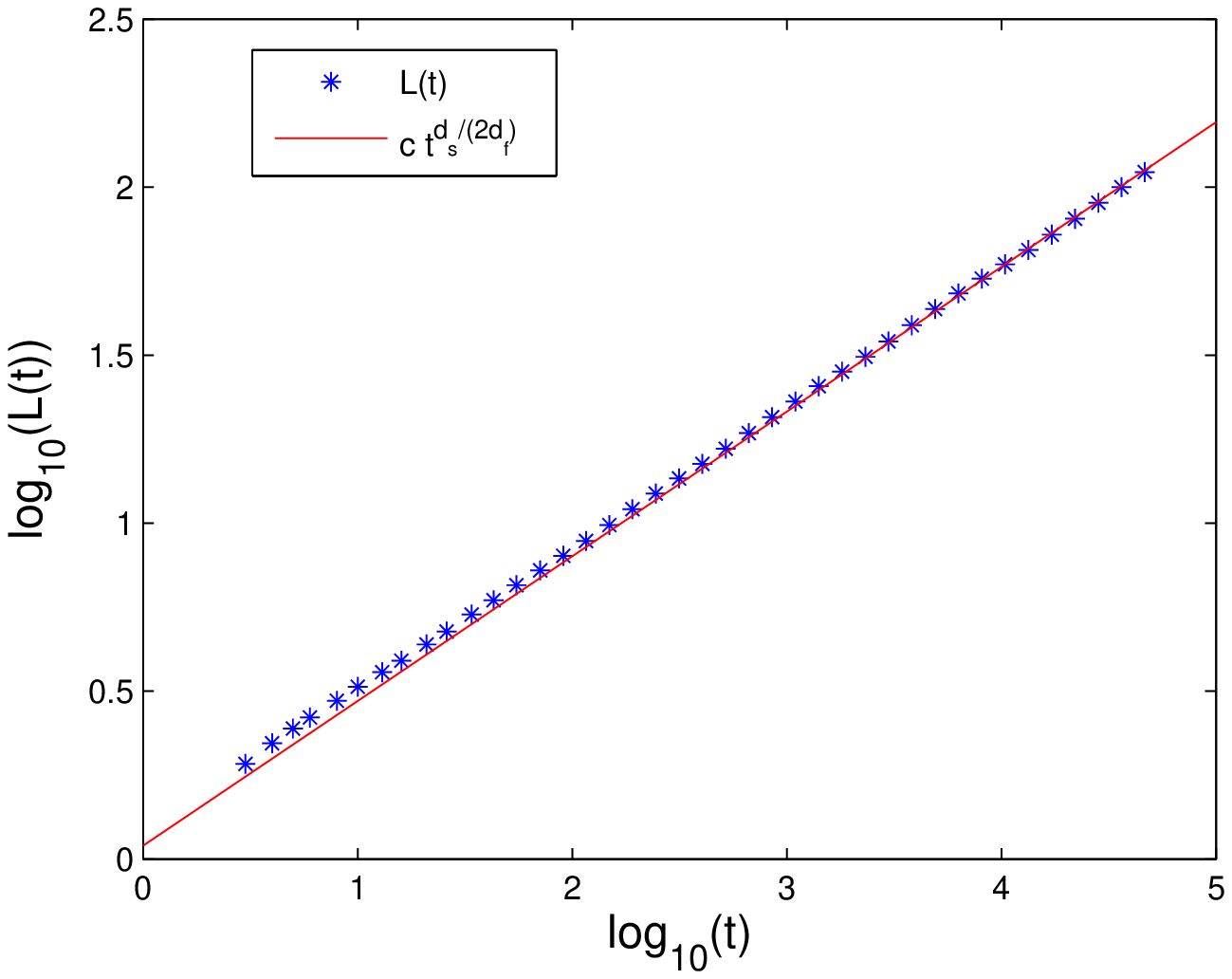}}} 
    \caption{Growth of $L(t)$ as a function of time for the $O(4)$ model on 
the  Sierpinski gasket, a comparison with the expected behaviour $c t^{d_s/(2 d_f)}$ is provided.}
\label{lcorr_O4}
\end{figure}

\begin{figure}
    \centering
   \rotatebox{0}{\resizebox{.5\textwidth}{!}{\includegraphics{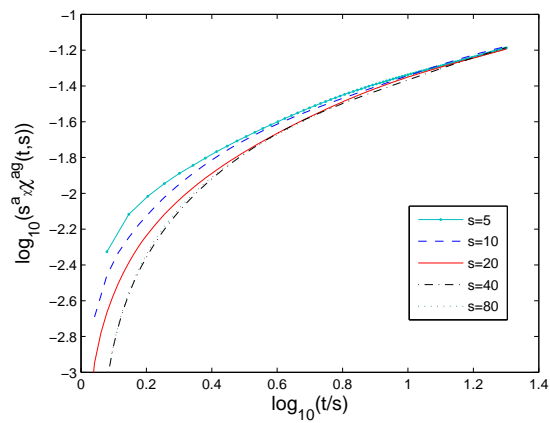}}} 
    \caption{Scaling of the susceptibility $\chi^{ag}(t,s)$, for large $t/s$ a good agreement with the scaling hypothesis (\ref{scalrespag}) with $a_{\chi}=d_s/2-1$ is evidenced.}
\label{scal_chi_O4}
\end{figure}

Up to now, by focusing on the case ${\cal N}=3$, we have shown that
the global pattern of behaviours observed on lattices is
maintained on the fractal structures we have considered,
both with $d_s<2$ and with $d_s>2$. Moreover, exponents such as $z$ and $a_\chi$ that
are known on lattices not to depend on ${\cal N}$, are found consistent
with the large-${\cal N}$ value. 
These results suggest that the property of being ${\cal N}$-independent 
may hold for any scaling system with ${\cal N}\geq 2$.
To check this hypothesis, we have considered the $O(4)$ model on the less 
computational demanding structure, i.e. the Sierpinski gasket. 
In Fig. \ref{lcorr_O4} we show the growth of the correlation length,
which is again in good agreement with an exponent $z=2d_f/d_s$.
Analogously, also $\chi ^{ag}(t,s)$, at large enough time ratios, 
scales as (\ref{scalrespag}) with 
an exponent very well consistent with the large-${\cal N}$ result
$a_{\chi}=d_s/2-1$, as shown in Fig. \ref{scal_chi_O4}.

\section{Conclusions} \label{concl}

In this paper, we have studied the 
phase-ordering kinetics of $O(3)$ and $O(4)$ models on self-similar 
physical graphs with known spectral dimension. 
On homogeneous lattices these models obey dynamical scaling.
We have shown by means of extensive numerical simulations 
that the same symmetry appears to be obeyed also 
on the networks considered.
The non-equilibrium exponents $z$ and $a_\chi$  
are consistent with the value provided by the  large-${\cal N}$ model. 
These results suggest that Equations (\ref{expa}) and (\ref{zinfty}) 
for $z$ and $a_\chi$ may hold for any ${\cal N} \geq 2$, 
for systems interested by dynamical scaling.
It would important to substantiate this hypothesis by means of analytic 
calculations that go beyond the large-${\cal N}$ limit by accessing
directly the case of finite ${\cal N}$. This could be 
possibly achieved by using a Gaussian Auxiliary Field 
approximation  on the equation of motion for the $O({\cal N})$ model \cite{desiena}.

On the other hand, concerning the issue of the generality of the scaling
property,
it would be interesting to study if the breakdown of dynamical scaling
always occurs in the $O(2)$ model on structures with $d_s<2$, similarly to 
what observed on homogeneous one-dimensional lattices \cite{rutenberg}.

\appendix
\section{$O(2)$ model} \label{O(2)}

\begin{figure}
    \centering
   \rotatebox{0}{\resizebox{.5\textwidth}{!}{\includegraphics{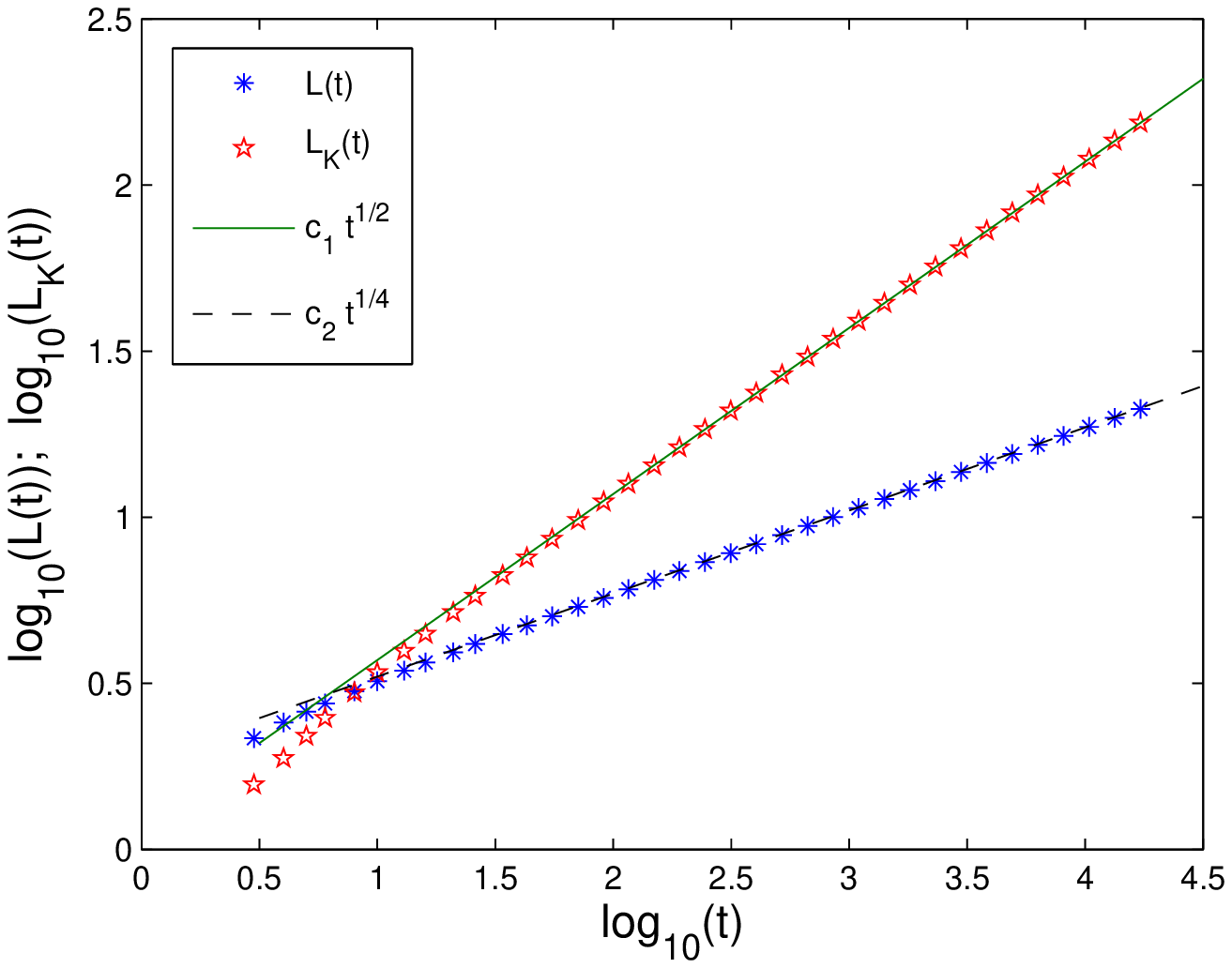}}}
    \caption{The growth of the correlation lengths $L(t)$ and $L_K(t)$.}
\label{O21D}
\end{figure}

In this appendix we present some results obtained for the $O(2)$ model 
using zero temperature heath bath dynamics (\ref{move}). First we discuss phase ordering 
in a one-dimensional system. We show that our dynamics reproduce the 
asymptotic behaviour predicted analytically in \cite{rutenberg} where 
zero-temperature phase ordering for the $O(2)$ model is studied by means 
of continuous time dissipative dynamics. 
In \cite{rutenberg} it is evidenced that the dynamical scaling is violated since
two different correlation lengths are present. The first one, 
the phase winding length $L(t)$,  is the characteristic length 
of the spin spin-spin correlation function defined by Eq. (\ref{corrlen}).
The other is the phase coherence length $L_K(t)$ representing the typical 
distance 
for which the winding of the spins along the system changes
its direction. In particular, calling $-\pi<\theta_i<\pi$ the angle between 
neighbouring
spins $\vec \sigma_i$ and $\vec \sigma_{i+1}$, one defines on each site the 
binary variable $k_i=\pm1$ if $\theta_i>0$ or $\theta_i<0$ respectively. 
According to \cite{rutenberg},  the
correlation function $G_K(r,t)=\langle k_i(t) k_{i+r}(t)\rangle$
scales differently from $G(r,t)$; in particular
\be
G(r,t)=g(r/L(t))=g(x)~~~~G_K(r,t)=g_k(r/L_K(t))=g_k(x)
\label{corrlen-rut}
\ee
with $L(t) \sim t^{1/4}$ and $L_K(t) \sim t^{1/2}$.
In our simulation with zero temperature heath bath dynamics we verify 
the laws (\ref{corrlen-rut}). In particular in Fig. \ref{O21D}
we display the growth in time of the two correlation function 
evidencing a very good agreement with the expected results.
We remark that adopting different dynamics, e.g. Metropolis, 
the analytically known behaviour is not easily reproduced in simulations because of
very-long transients 
(only the asymptotic behaviour is expected to be independent of the transition
rates).

\begin{figure}
    \centering
   \rotatebox{0}{\resizebox{.5\textwidth}{!}{\includegraphics{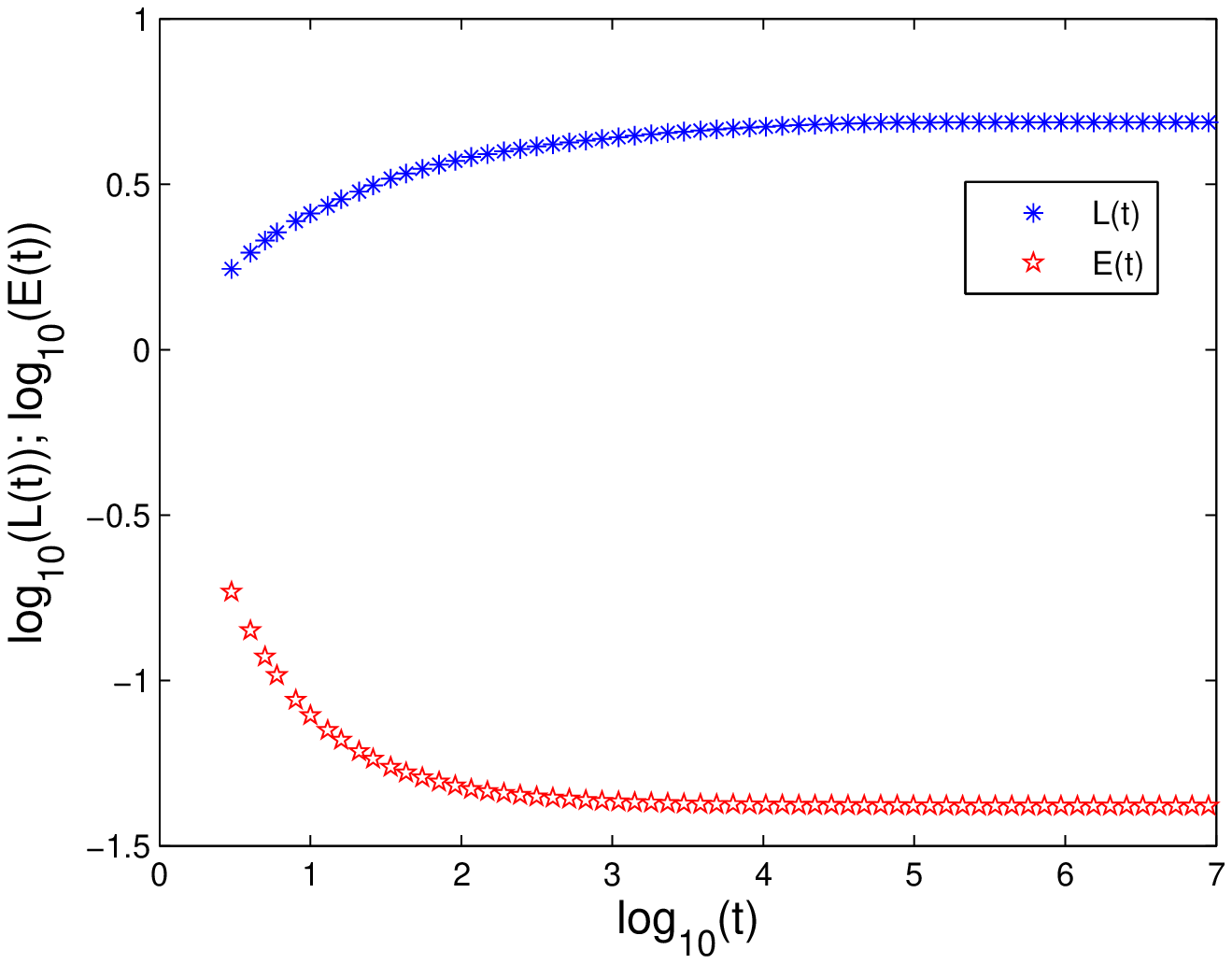}}}
    \caption{Correlation lengths $L(t)$ and energy $E(t)$ as a function of time 
      in the $O(2)$ model on the Sierpinski gasket fractal.
      $L(t)$ is defined by Eq. \ref{corrlen}. The energy scale is set so 
      that the energy of the ground state is zero.}
\label{O2gas}
\end{figure}

As explained in Sec. \ref{model} we have initially considered the $O(2)$ 
rotator model on fractals, and we obtain 
that the system gets frozen into metastable 
states. In particular  the energy $E(t)$ does not tend to zero and the correlation 
length does not diverge. Figure \ref{O2gas} shows this 
behaviour for the Sierpinski gasket. Analogous behaviours are present
on other fractals.

\end{document}